\documentclass[aip,jap,reprint,superscriptaddress,amsmath, showpacs]{revtex4-1}
\usepackage{graphicx, dcolumn}
\begin{document}
\title{Exchange bias effect in martensitic epitaxial Ni-Mn-Sn thin films applied to pin CoFeB/MgO/CoFeB magnetic tunnel junctions}
\author{N. Teichert}\email{nteichert@physik.uni-bielefeld.de}
\affiliation{Center for Spinelectronic Materials and Devices, Department of Physics, Bielefeld University, 33615 Bielefeld, Germany}

\author{A. Boehnke}
\affiliation{Center for Spinelectronic Materials and Devices, Department of Physics, Bielefeld University, 33615 Bielefeld, Germany}

\author{A. Behler}
\affiliation{IFW Dresden, Institute for Complex Materials, P.O. Box 27 01 16, 01171 Dresden, Germany}

\author{B. Weise}
\affiliation{IFW Dresden, Institute for Complex Materials, P.O. Box 27 01 16, 01171 Dresden, Germany}

\author{A. Waske}
\affiliation{IFW Dresden, Institute for Complex Materials, P.O. Box 27 01 16, 01171 Dresden, Germany}

\author{A. H\"utten} 
\affiliation{Center for Spinelectronic Materials and Devices, Department of Physics, Bielefeld University, 33615 Bielefeld, Germany}

\date{\today}
\begin{abstract}
The exchange bias effect is commonly used to shift the coercive field of a ferromagnet. This technique is crucial for the use of magnetic tunnel junctions as logic or memory devices. Therefore, an independent switching of the two ferromagnetic electrodes is necessary to guarantee a reliable readout. Here, we demonstrate that the intrinsic exchange bias effect of Ni-Mn-Sn can be used to apply a unidirectional anisotropy to magnetic tunnel junctions. For this, we use epitaxial Ni-Mn-Sn films as pinning layers for microfabricated CoFeB/MgO/CoFeB magnetic tunnel junctions. We compare the exchange bias field ($H_{\text{EB}}$) measured after field cooling in $-10$\,kOe external field by magnetization measurements with $H_{\text{EB}}$ obtained from tunnel magnetoresistance measurements. Consistent for both methods we find an exchange bias of about $H_{\text{EB}}=130$\,Oe at 10\,K, which decreases with increasing temperature and vanishes above 70\,K.
\end{abstract}
\maketitle
The exchange bias effect (EB) describes a unidirectional magnetic anisotropy resulting in a shift of the magnetic hysteresis along the direction of the applied field.\cite{Meiklejohn1956} It is observed in structures with interfaces between ferromagnetic (FM) and antiferromagnetic (AF) phases, e.g. thin film structures with FM and AF layers.\cite{Nogues1999} EB is commonly used to pin magnetic electrodes in magnetic tunnel junctions (MTJs).  The magnetic shape memory Heusler compounds Ni-Mn-X (X= Sn, In, Sb) show an intrinsic EB in the martensitic state at low temperature caused by FM and AF regions in the material.\cite{Khan2007, Khan2007apl, Pathak2009, Machavarapu2013} Accordingly, these compounds are promising candidates for pinning ferromagnetic electrodes of MTJs without the commonly used antiferromagnets, MnIr and MnPt.\cite{Kaemmerer2004, Lee2006, Teixeira2012}
Replacing these materials in industrial applications is desirable because of the rarity and high cost of iridium and platinum.
Here, we present how Ni-Mn-Sn thin films can serve as pinning layer in CoFeB/MgO/CoFeB MTJs and thus, show the technological applicability of the EB of Ni-Mn-X Heusler compounds. Since Ni-Mn-Sn is magnetic and shows an intrinsic EB, in principle it could serve as pinning layer and magnetic electrode at once. Nevertheless, we use CoFeB electrodes because they are well established\cite{Ikeda2008} and the spin-polarization of Ni-Mn-Sn is small, which is unfavorable for electrodes in MTJs.\cite{Ye2010}

\begin{figure}
\centering
\includegraphics[width=7.54 cm]{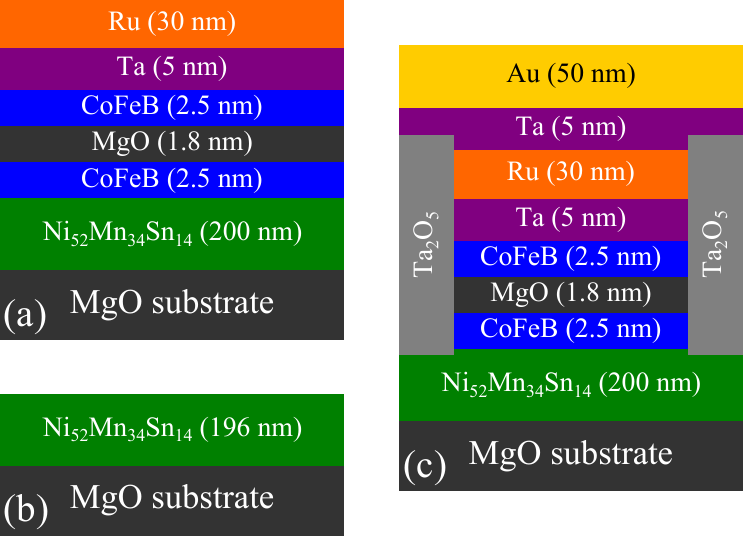}
\caption{\label{fig1} Sketches of the layer system: (a) as deposited (sample A), (b) etched (sample B), and (c) after annealing and nanofabrication (sample C).}
\end{figure}

The samples were fabricated by magnetron sputtering and subsequent e-beam lithography. In a first step we deposit the 200\,nm thick Ni$_{52}$Mn$_{34}$Sn$_{14}$ layer on MgO(001) substrates by co-sputtering from elemental targets as described in Ref.~\onlinecite{Auge2012}. The substrate temperature during the deposition was 650$^{\circ}$C. After the deposition process the samples are cooled to room temperature and the additional layers Co$_{\text{40}}$Fe$_{\text{40}}$B$_{\text{20}}$/MgO/Co$_{\text{40}}$Fe$_{\text{40}}$B$_{\text{20}}$/Ta/Ru were deposited with film thicknesses as shown in Fig.~\ref{fig1}(a). This "as deposited" sample is referred to as sample A.

So as to get a single Ni$_{52}$Mn$_{34}$Sn$_{14}$ layer as reference sample the upper layers of sample A were removed by Ar ion beam etching while the progress is monitored by secondary ion mass spectrometry. Hereby, we obtain a 196\,nm thick Ni-Mn-Sn layer on MgO substrate denoted as sample B (cf. Fig.~\ref{fig1}(b)). The film thickness of Ni-Mn-Sn was checked by X-ray reflectometry measurements.

In order to prepare the MTJs for tunnel magnetoresistance (TMR) measurements sample A is annealed at 350$^{\circ}$C for one hour for proper crystallization of the CoFeB electrodes.\cite{Schmalhorst2007} Afterwards ellipsoidal MTJs (300\,nm x 180\,nm) were patterned out of the layer stack by e-beam lithography and subsequent Ar ion milling.  The ion milling was stopped right below the lower CoFeB layer in order to keep the Ni-Mn-Sn film intact. This is necessary because the crystallite size of shape memory materials has a high impact on the transformation characteristics and the martensitic transformation can be impeded in too small crystals.\cite{Meng2002, Malygin2009, Teichert2015}
The MTJs were insulated by Ta$_{2}$O$_{5}$ and equipped with Ta/Au contact pads by RF and DC magnetron sputtering. This annealed and patterned sample is referred to as sample C and the final layer structure is sketched in Fig.~\ref{fig1}(c).

Composition and crystal structure of the Ni-Mn-Sn layer were examined by energy dispersive X-ray spectroscopy and X-ray diffraction measurements, respectively. We observe crystallization in $L2_1$ structure and epitaxial growth with the relation MgO(001)[110]$\vert\vert$Ni-Mn-Sn(001)[100].\cite{S_XRD}
The magnetization was studied using a vibrating sample magnetometer (Quantum Design PPMS) with in-plane applied magnetic field. TMR was measured using common 2-probe method with a constant 100\,mV DC bias voltage and a Cryogenic He cryostat system where the external field is applied along the major axis of the ellipsoidal MTJs. The resistance-area product of the tunnel junctions in parallel configuration is RA$=3.0(6)$\,k$\Omega$\,$\mu$m$^2$, which is in good agreement with other reports on CoFeB/MgO/CoFeB MTJs.\cite{Lee2006} Ni-Mn-Sn is known to show sizable resistance changes in dependence of temperature and magnetic field.\cite{Singh2011, Yuzuak2013, Teichert2015} However, those are part of the lead resistance, which is three to four orders of magnitude smaller than the junction resistance, and thus, can be neglected.

\begin{figure}
\centering
\includegraphics[width=8.5 cm]{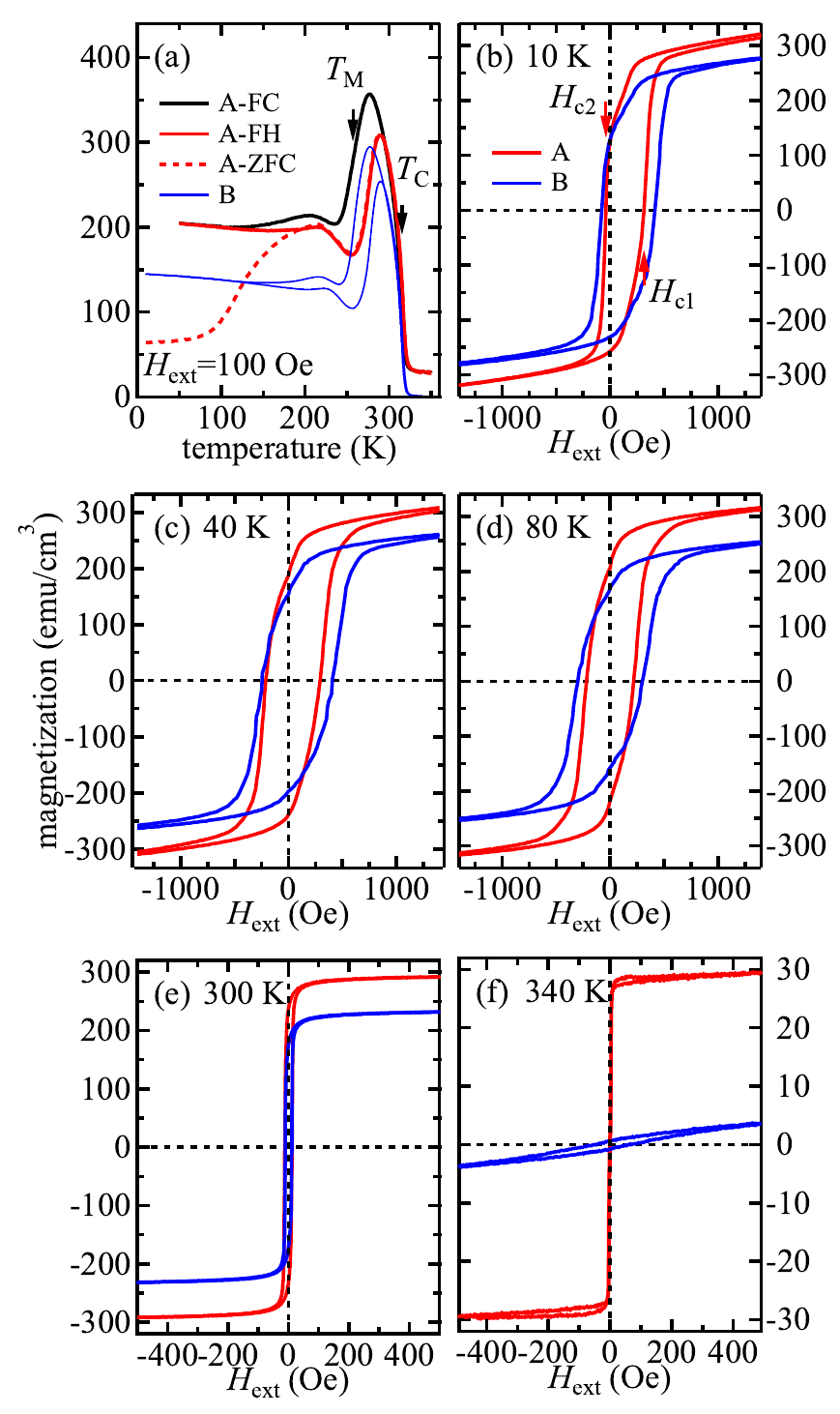}
\caption{\label{fig2}Magnetization measurements of samples A and B (cf. Fig.~\ref{fig1}(a) and Fig.~\ref{fig1}(b): (a) $M(T)$ curves at 100\,Oe applied field. The FC (solid black line) and FH (solid red line) curves of sample A envelop a thermal hysteresis due to the martensitic transition. The FH and ZFC (dashed red line) curves are split at low temperature. The martensitic transformation temperature $T_{\text{M}}$ and Curie temperature $T_{\text{C}}$ of Ni-Mn-Sn are indicated by arrows. FC and FH curves of sample B are shown as blue lines. (b)-(f) $M(H)$ curves after field cooling at -10\,kOe of samples A (red line) and B (blue line). The EB leads to a shift of $H_{\text{c1}}$ and $H_{\text{c2}}$ in positive field direction at low temperature.}
\end{figure}

In order to determine the magnetic properties and martensitic transformation of the Ni-Mn-Sn the temperature dependence of the magnetization was studied. The magnetization values given in the paper are the measured magnetic moments normalized with the total volume of magnetic material (205\,nm thickness for sample A and 196\,nm for sample B). Fig.~\ref{fig2}(a) shows the magnetization versus temperature for samples A and B in field cooling (FC) and field heating (FH), and (for sample A) during heating after cooling the sample in zero magnetic field (ZFC) measured in low external field. Apart from the overall higher magnetization of sample A due to the CoFeB layers, samples A and B show the same temperature dependence. The distinct drop of the magnetization upon cooling results from the martensitic transformation. For Ni-Mn-Sn, the magnetization of martensite is lower than that of austenite. The reason for the magnetization change is a change in the alignment of magnetic moments of the Mn atoms on Mn sites (Mn$_{\text{1}}$) and Mn atoms on Sn sites (Mn$_{\text{2}}$), which are known to couple antiferromagnetically in austenite and martensite. However, due to changed lattice constants the antiferromagnetic coupling between Mn$_{\text{1}}$ and Mn$_{\text{2}}$ is strengthened in the martensite.\cite{Aksoy2009, Sokolovskiy2012, Behler2013}
The martensitic transformation temperature and Curie temperature of the Ni-Mn-Sn are $T_{\text{M}}=257\,$K and $T_{\text{C}}=316\,$K, respectively, determined from inflection points of the FC magnetization curve. Above $T_{\text{C}}$ the magnetization of sample A drops to 27\,emu\,cm$^{-3}$. This remaining magnetization results from the CoFeB, which has a higher Curie temperature. The splitting between the ZFC and FC curve originates from the coexistence of ferromagnetic and antiferromagnetic regions at low temperatures in the martensite phase, which is the basis for the EB.\cite{Krenke2005,Wang2011}

Commonly, the intrinsic EB in Ni-Mn-Sn is determined via isothermal magnetization curves at low temperature after cooling the specimen in an external magnetic field. Fig.~\ref{fig2}(b)-(f) depict the magnetic hysteresis loops of samples A (red lines) and B (blue lines) after field cooling to 10\,K in -10\,kOe external field. From Fig.~\ref{fig2}(b) and (c) it is clearly visible that at low temperatures the magnetic hysteresis is shifted and the positive coercive field $H_{\text{c1}}$ is larger than the negative coercive field $H_{\text{c2}}$. The exchange bias field is defined as $H_{\text{EB}}=(H_{\text{c1}}+H_{\text{c2}})/2$. Above 80\,K (Fig.~\ref{fig2}(d)-(e) the curves are symmetric and $H_{\text{EB}}=0$. 
Above $T_{\text{C}}$ (cf. Fig.~\ref{fig2}(f)) the magnetization of sample B is small and sample A mainly shows the magnetization of CoFeB. The saturation magnetization ($M_\text{S}$) of CoFeB appears too small, because the magnetic moment is normalized using a film thickness of 205\,nm as described above. Considering only 5\,nm leads to $M_\text{S}=1150$\,emu\,cm$^{-3}$ for CoFeB, which coincides well with the literature.\cite{Ikeda2010}

Sample A shows aside from higher magnetization also lower coercive fields and lower $H_{\text{EB}}$ than sample B caused by the presence of CoFeB, which itself has low coercivity as seen in Fig.~\ref{fig2}. However, from the $M(H)$ curves in Fig.~\ref{fig2} separate magnetic switching of one or both CoFeB layers is not observed. Thus, all magnetic layers are ferromagnetically coupled in the unpatterned sample A: the lower CoFeB layer by the direct contact to the Ni-Mn-Sn and the upper layer most likely through pinholes in the MgO barrier.

\begin{figure}
\centering
\includegraphics[width=8.5 cm]{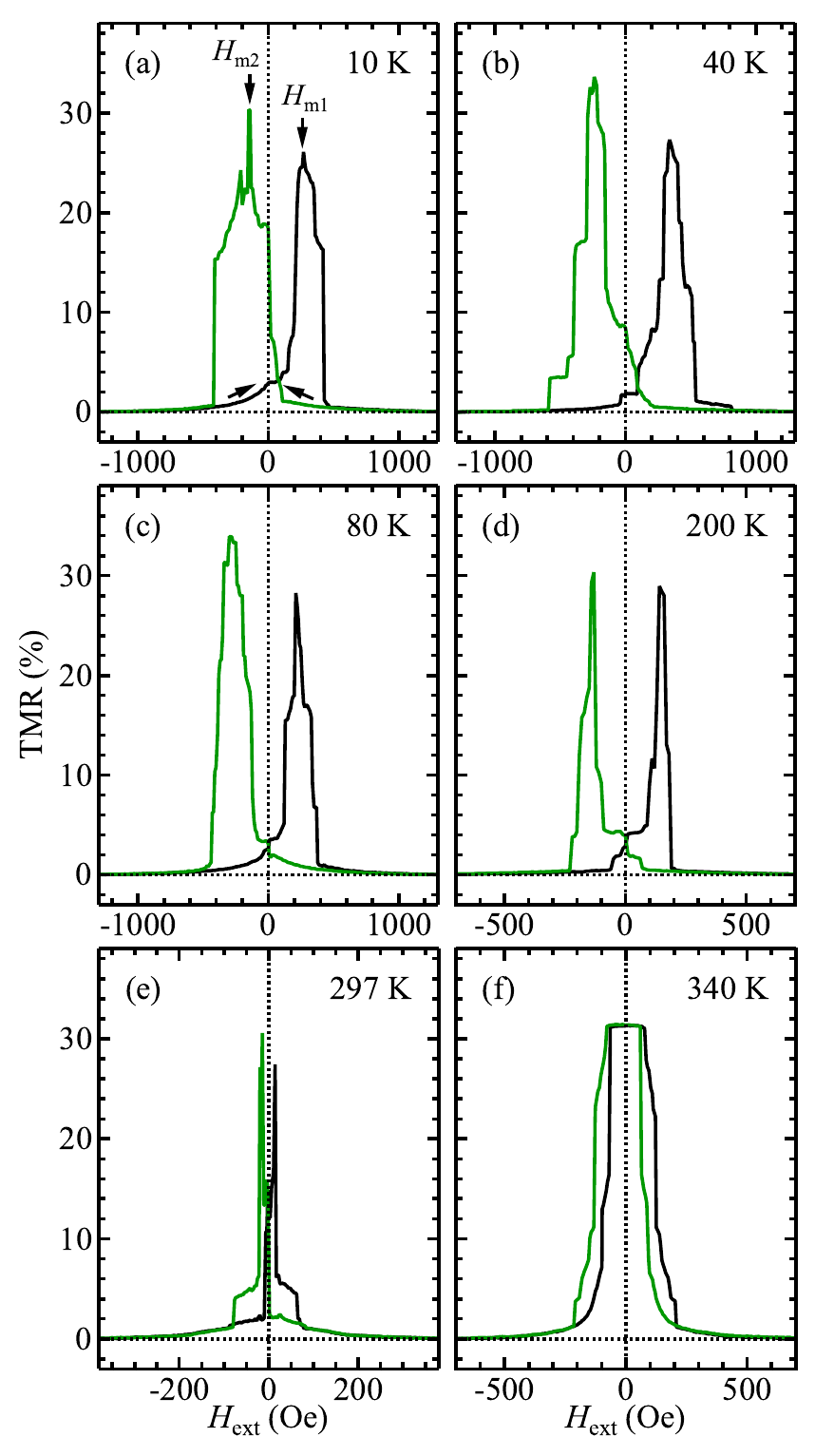}
\caption{\label{fig3} TMR curves measured using sample C at 100\,mV bias voltage after field cooling at -10\,kOe. Black and green lines are measured under increasing and decreasing field, respectively.  At low temperature the EB is visible ((a) and (b)). (f) shows the TMR for paramagnetic Ni-Mn-Sn where the TMR curve is symmetric.}
\end{figure}

In the following, the EB is determined from TMR measurements using sample C. Fig.~\ref{fig3} shows the TMR ratio defined as $\text{TMR}=(R-R_{\text{p}})/R_{\text{p}}$ versus external field. $R_{\text{p}}$ denotes the resistance at parallel alignment of both CoFeB electrodes, i.e. the resistance at high applied field. Hereby, the EB is defined as $H_{\text{EB}}^{\text{TMR}}=(H_{\text{m1}}+H_{\text{m2}})/2$ and deduced from $H_{\text{m1}}$ and $H_{\text{m2}}$, which denote the external fields of maximum TMR in increasing and decreasing field, respectively.
The exchange bias effect at low temperature is clearly visible from Fig.~\ref{fig3}(a) and (b) since $H_{\text{m1}}>\vert H_{\text{m2}}\vert$ and vanishes above 80\,K (Fig.~\ref{fig3}(c)-(f))
in analogy to Fig.~\ref{fig2}(b) and (f).

Below $T_{\text{C}}$ (Fig.~\ref{fig3}(a)-(e)) the curves are asymmetric and show spikes of maximum TMR with different amplitudes under increasing and decreasing external field. Furthermore, the TMR amplitude fluctuates around 30\% and does not show a systematic temperature dependence. This is caused by the absence of a fully antiparallel state of the magnetic electrodes and is contrary to the situation above $T_{\text{C}}$ (Fig.~\ref{fig3}(f)) where symmetric curves with a plateau of maximum TMR are observed.
The resistivity change occurs in multiple steps, which reflect Barkhausen jumps of domain walls in the upper CoFeB electrode and the lower electrode, which is a Ni-Mn-Sn/CoFeB bilayer below $T_{\text{C}}$ and CoFeB above $T_{\text{C}}$. 
Since the Ni-Mn-Sn layer was left intact during the patterning process the magnetic domains go beyond the MTJs and are not pinned by their geometry. This causes a random influence on the shape of the TMR curves below $T_{\text{C}}$. Barkhausen jumps are not observed in the $M(H)$ curves in Fig.~\ref{fig2} because the relevant sample area (several square millimeters) is much larger than the domain size, which is apparently not the case for the MTJs.

The plateau of maximum TMR at 340\,K around $H_{\text{ext}}=0$ is caused by AF coupling of the magnetic electrodes. This can be explained by magnetostatic coupling of the CoFeB layers induced by uncompensated magnetic poles at the edges of the electrodes.\cite{Anguelouch2000}
Below $T_{\text{C}}$ at zero applied field the alignment of the magnetic electrodes is only partly AF. The remanence of Ni-Mn-Sn is smaller than its saturation magnetization (cf. Fig.~\ref{fig2}(b)-(f)) and the AF coupling of the CoFeB electrodes is apparently to weak to fully align the magnetization of Ni-Mn-Sn below.\cite{S_XRD}

The quite modest TMR amplitudes of maximum 34\% despite our choice of CoFeB as magnetic electrode material and 1.8\,nm MgO barrier are probably caused by rather poor quality of the tunnel barrier. The surface roughness of the underlying Ni-Mn-Sn layer is with $r\approx 1\,$nm (measured by atomic force microscopy using a separate Ni-Mn-Sn film, not shown) larger than of conventional Ta underlayers ($r\approx 0.2\,$nm\,\cite{Hayakawa2005}) which reduces the quality of the MgO barrier, and hence, the TMR amplitude.
Also the MTJ layer stack is not optimized by means of film thicknesses and selection of the optimum materials for the magnetic electrodes and the tunnel barrier in order to obtain maximum TMR.

\begin{figure}
\centering
\includegraphics[width=8.5 cm]{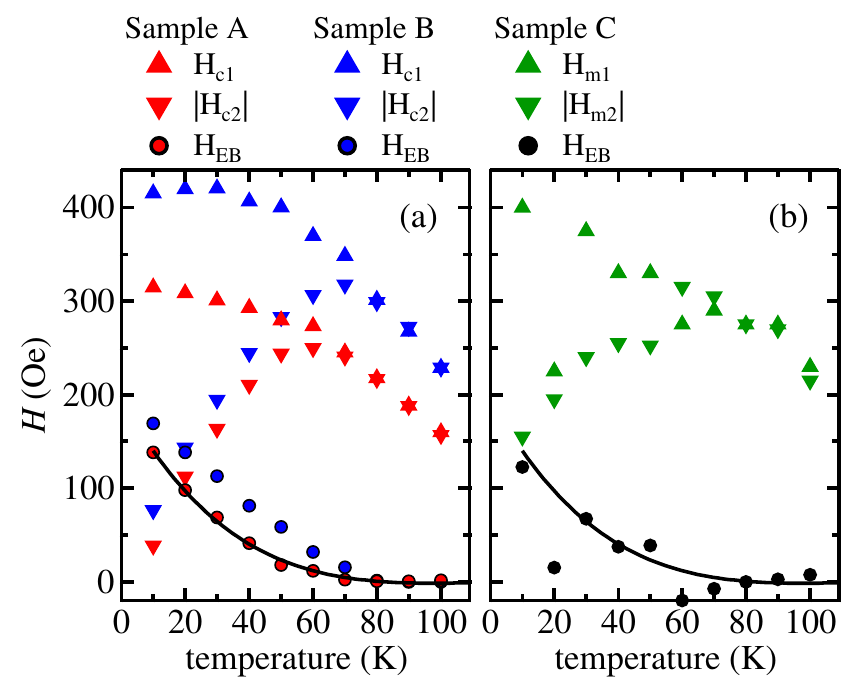}
\caption{\label{fig4} Temperature dependence of the EB after field cooling at -10\,kOe. (a) $H_{\text{c1}}$, $H_{\text{c2}}$, and $H_{\text{EB}}$ determined from magnetization measurements of samples A (red symbols) and B (blue symbols). The black line is a guide to the eye and identical in (a) and (b). (b) $H_{\text{m1}}$, $H_{\text{m2}}$, and $H_{\text{EB}}$ determined from TMR curves using sample C.}
\end{figure}

The temperature dependence of the EB is depicted in Fig.~\ref{fig4} where both methods of determining EB are compared. Fig.~\ref{fig4}(a) shows $H_{\text{c1}}$, $H_{\text{c2}}$, and $H_{\text{EB}}$ as determined by magnetization measurements of samples A (red symbols) and B (blue symbols). At low temperature $\vert H_{\text{c1}}\vert$ and $\vert H_{\text{c2}}\vert$ are different. With increasing temperature $\vert H_{\text{c2}}\vert$ strongly increases while $\vert H_{\text{c1}}\vert$ mildly decreases up to 60\,K. Above 70\,K the values of positive and negative coercive fields are equal and decrease with increasing temperature. Accordingly, $H_{\text{EB}}$ strongly decreases with increasing temperature from $H_{\text{EB}}=138\,$Oe for sample A and $H_{\text{EB}}=169\,$Oe for sample B at 10\,K to zero above 70\,K. The magnitude and temperature dependence of $H_{\text{EB}}$ of sample B is comparable to bulk Ni-Mn-Sn.\cite{Khan2007,Li2007} 
Both coercive fields as well as $H_{\text{EB}}$ of sample A are reduced by the CoFeB layers.

Fig.~\ref{fig4}(b) depicts the corresponding results from TMR measurements. The depicted data are average values obtained from TMR curves of two different MTJs measured during the first and second field loop at each temperature point.
$H_{\text{m1}}$ and $H_{\text{m2}}$ show the same trends as $H_{\text{c1}}$ and $H_{\text{c2}}$. As mentioned earlier, below $T_{\text{C}}$ the measured TMR curves exhibit Barkhausen noise. This acts as a random influence on $H_{\text{m1}}$ and $H_{\text{m2}}$ and can be of the same order of magnitude as the exchange bias which leads to a very small $H_{\text{EB}}$ at 20\,K and a negative $H_{\text{EB}}$ at 60\,K. Nevertheless, it is clearly visible from Fig.~\ref{fig4}(a) and (b) that $H_{\text{EB}}$ has the same magnitude and temperature dependence for the samples A and C. So, the intrinsic EB of Ni-Mn-Sn can be observed in TMR measurements.
 
In summary, we investigated the exchange bias effect of MgO(substrate)/Ni-Mn-Sn/CoFeB/MgO/CoFeB and MgO(substrate)/Ni-Mn-Sn thin film structures after field cooling by two different methods: direct magnetization measurements and TMR measurements. Magnetization measurements are used to quantify the intrinsic EB of the Ni-Mn-Sn layer and the influence of the magnetic moment of the thin CoFeB layers. TMR measurements are sensitive to the interaction between the Ni-Mn-Sn layer and the CoFeB tunnel electrodes. Since we have shown a comparable EB effect in MTJs and magnetization measurements, we conclude that epitaxial Ni-Mn-Sn thin films can serve as pinning layers in these devices.

\begin{acknowledgments}
The authors gratefully acknowledge funding by the Deutsche Forschungsgemeinschaft through SPP 1599 ``Ferroic Cooling''. Collaboration with the European-Japanese FP7 project HARFIR is gratefully acknowledged. We also thank J.-M. Schmalhorst for fruitful discussion and D. Kucza for AFM measurements.
\end{acknowledgments}

\end{document}